\documentstyle[preprint,aps,epsf]{revtex}

\RequirePackage{array}
\def\singlespace {\smallskipamount=3.75pt plus1pt minus1pt
                  \medskipamount=7.5pt plus2pt minus2pt
                  \bigskipamount=15pt plus4pt minus4pt
                  \normalbaselineskip=12pt plus0pt minus0pt
                  \normallineskip=1pt
                  \normallineskiplimit=0pt
                  \jot=3.75pt
                  {\def\smallskip {\vskip\smallskipamount}}
                  {\def\medskip   {\vskip\medskipamount}}
                  {\def\bigskip   {\vskip\bigskipamount}}
                  {\setbox\strutbox=\hbox{\vrule
                    height10.5pt depth4.5pt width 0pt}}
                  \parskip 7.5pt
                  \normalbaselines}
\def\middlespace {\smallskipamount=5.625pt plus1.5pt minus1.5pt
                  \medskipamount=11.25pt plus3pt minus3pt
                  \bigskipamount=22.5pt plus6pt minus6pt
                  \normalbaselineskip=22.5pt plus0pt minus0pt
                  \normallineskip=1pt
                  \normallineskiplimit=0pt
                  \jot=5.625pt
                  {\def\smallskip {\vskip\smallskipamount}}
                  {\def\medskip   {\vskip\medskipamount}}
                  {\def\bigskip   {\vskip\bigskipamount}}
                  {\setbox\strutbox=\hbox{\vrule
                    height15.75pt depth6.75pt width 0pt}}
                  \parskip 11.25pt
                  \normalbaselines}
\def\doublespace {\smallskipamount=7.5pt plus2pt minus2pt
                  \medskipamount=15pt plus4pt minus4pt
                  \bigskipamount=30pt plus8pt minus8pt
                  \normalbaselineskip=30pt plus0pt minus0pt
                  \normallineskip=2pt
                  \normallineskiplimit=0pt
                  \jot=7.5pt
                  {\def\smallskip {\vskip\smallskipamount}}
                  {\def\medskip   {\vskip\medskipamount}}
                  {\def\bigskip   {\vskip\bigskipamount}}
                  {\setbox\strutbox=\hbox{\vrule
                    height21.0pt depth9.0pt width 0pt}}
                  \parskip 15.0pt
                  \normalbaselines}

\def\ga{\gamma}

\def\frac#1#2{\textstyle{{{#1} \over {#2}}}}

\def\lsim{\mathrel{\rlap{\lower4pt\hbox{\hskip1pt$\sim$}}
    \raise1pt\hbox{$<$}}}
\def\gsim{\mathrel{\rlap{\lower4pt\hbox{\hskip1pt$\sim$}}
    \raise1pt\hbox{$>$}}}

\newcommand{\beq}{\begin{equation}}
\newcommand{\eeq}{\end{equation}}
\newcommand{\bea}{\begin{eqnarray}}
\newcommand{\eea}{\end{eqnarray}}

\tightenlines
\begin{document}
\preprint{
\hfill$\vcenter{\hbox{\bf IUHET-483} \hbox{July
             2005}}$  }

\title{\vspace*{.75in}
Signals for Low Scale Gravity in the Process $\gamma \gamma \to ZZ$}

\author{Micheal S. Berger
\footnote{Electronic address: berger@indiana.edu} and 
Brandon Zerbe\footnote{Electronic address: bzerbe@indiana.edu}}

\address{
Physics Department, Indiana University, Bloomington, IN 47405, USA}

\maketitle

\thispagestyle{empty}

\begin{abstract}
We investigate the sensitivity of future photon-photon colliders to low scale 
gravity scenarios via the process $\gamma\gamma \to ZZ$ where 
the Kaluza-Klein boson exchange contributes only when the initial state photons
have opposite helicity. We contrast this with the situation for the process
$\gamma \gamma \to \gamma \gamma$ where the $t$ and $u$ channel also 
contribute. We include the one-loop Standard Model background whose 
interference with the graviton exchange determines the experimental 
reach in measuring any deviation from the Standard Model expectations
and explore how polarization can be exploited to enhance the signal
over background. We  find that a 1~TeV linear collider has an experimental 
reach to mass scale of about 4~TeV in this channel.
\end{abstract}


\newpage

\section{Introduction}
In the last few years the most popular speculative idea in 
theoretical particle physics has been the possibility that 
extra spacetime dimensions exist. Much of the interest in this 
area was stimulated by the 
realization that constraints on the extra dimensions were relatively mild
if only gravity and not the Standard Model gauge interactions was able to 
propagate in the extra dimensions or 
bulk\cite{Arkani-Hamed:1998rs,Antoniadis:1998ig}. This led to 
the possibility that
the effective Planck scale in the extra dimensions was much lower than 
the commonly used four-dimensional Planck scale. If the effective Planck scale
is of order a few TeV, then speculation arose that extra dimensions might 
help resolve the hierarchy problem and the electroweak scale effects of the 
extra dimensions might appear in future collider experiments. Gauss's law
links the value of the effective Planck scale in the bulk to the conventional
Planck scale via
\begin{eqnarray}
M_{pl}^2\sim R^nM_S^{n+2}\;.
\label{scales}
\end{eqnarray}

Physical effects can present themselves 
via graviton exchange
at future colliders, and an interesting
class of processes are the pair production of gauge bosons in the 
photon-photon collider.
The process $\gamma \gamma \to \gamma \gamma$ has been studied 
before\cite{Cheung:1999ja,Davoudiasl:1999di,Choudhury:1999gp}. The processes 
$\gamma \gamma \to W^+W^-$ and $\gamma \gamma \to ZZ$ were studied in 
Ref.~\cite{Rizzo:1999sy}. In the latter process the Standard Model 
contribution 
$\gamma \gamma \to ZZ$ is known\cite{Jikia:1993di,Berger:1993tr} but was not 
included. The process $\gamma \gamma \to ZZ$ is particularly
attractive for the following reasons: (1) it provides another channel
with which to assess the universality of the gravitational couplings 
to the gauge bosons; (2) the angular dependence of $\gamma \gamma \to ZZ$ is 
different from $\gamma \gamma \to \gamma \gamma$ because it occurs only through
the $s$-channel while $\gamma \gamma \to \gamma \gamma$ occurs through the
$s$, $t$, and $u$ channels; (3) since only the $s$-channel contributes to 
Kaluza-Klein (KK) process of $\gamma \gamma \to ZZ$ 
and the KK state is spin-two, 
we find the only helicity amplitudes which do not vanish have opposite initial 
photon helicities; (4) the $Z$ boson's transverse and 
longitudinal polarizations 
can be exploited by measuring the angular distribution of its decay products. 

Our emphasis here will be on the particular process $\ga\ga\to ZZ$ for which
the complete calculation including the full Standard Model background has not 
been performed\footnote{After this work was completed, we became aware of a 
paper\cite{Choudhury:2002zm} which included an approximate calculation of the 
Standard Model background and calculated the helicity amplitudes. Apart from 
some obvious typographical errors, we agree with the angular dependences of 
their helicity amplitudes and obtain 
similar numerical results. We have in addition 
included the photon-photon luminosity and explored the role of polarization 
in isolating the signal, and have derived bounds on the scale $M_S$.}.
We also present the helicity amplitudes for 
$\ga\ga\to\ga\ga$ which provide a basis for comparison and also allow us to 
make particular points about the properties of these processes that can be
exploited in a comprehensive analysis of all the final states. 

The Standard Model helicity amplitudes for $\gamma \gamma \to ZZ$ were
first published in Ref.~\cite{Jikia:1993di} and their analytic form was 
confirmed shortly thereafter\cite{Berger:1993tr}. Numerical calculations of 
the cross sections were also performed in 
Refs.~\cite{Bajc:1993hp,Dicus:1993ux}. More recently the helicity amplitudes
were again derived as a background for a search for possible virtual 
supersymmetric particles contribution to the loop 
diagrams\cite{Gounaris:1999hb}. The three calculations for the analytic 
expressions for the matrix elements show complete agreement (apart from
a typo in Ref.~\cite{Jikia:1993di} explained in 
Refs.~\cite{Berger:1993tr,Gounaris:1999hb}, and taking into 
account an unconventional definition of the Mandelstam 
variables $t$ and $u$ used 
in Ref.~\cite{Jikia:1993di}). 
The fermion loop contribution in the Standard Model was first 
calculated\cite{Glover:1988rg} in
the context of the gluon fusion process $gg\to ZZ$.
The results for that process 
are easily adapted 
to the process considered $\ga\ga\to ZZ$ considered here.

At high energies
where the low scale gravity signal should be most prominent, the Standard 
Model cross sections are dominated by the $W$ loop diagrams (as one 
expects since the $W$ boson is spin-one).
Numerically at energies sufficiently far above threshold the 
cross section for the background of $\ga\ga\to ZZ$ is an 
order of magnitude larger 
than than the background of $\ga\ga\to\ga\ga$. This can be 
understood simply by comparing the 
size of the $WWZ$ coupling to the $WW\ga$ where the ratio is determined 
solely by the Weinberg angle.

Photon beams can be realized at a future $e^+e^-$ collider by Compton 
backscattering laser beams off the electron or positron 
beam\cite{Ginzburg:1981vm,Ginzburg:1982yr,Telnov:1989sd}.
By exploiting circular polarization of the lasers and polarizing the 
electron beams, the contribution to cross sections from various initial state
photon helicities can be adjusted. 
 
We have obtained the contributions for the graviton exchange signal for 
both $\gamma \gamma \to \ga\ga$ and $\gamma \gamma \to ZZ$ at the helicity 
amplitude level through the use of FORM\cite{Vermaseren:2000nd}. 
If the photon-photon option at a next generation linear
collider becomes a real possibility in the future, this will facilitate 
detailed investigations of these processes putting in the full inteference
with the Standard Model contributions and retaining all information on the 
polarization of the incident photon beams. Furthermore for the $ZZ$ final 
state, more sophisticated cuts on the $Z$ boson decay products via the 
density matrix formalism can be exploited to improve sensitivity to any 
signal. Finally having the helicity amplitudes at our disposal allows us to 
understand angular distributions that reflect
the fact that graviton exchange is spin-two in nature.

Other processes have been considered as probes of low scale gravity. For 
cases where gravitons appear as virtual particles, calculations have been 
performed for the production of fermions\cite{Mathews:1998kf}, 
gauge bosons\cite{Atwood:1999cy,Agashe:1999qp}, 
Higgs bosons\cite{Rizzo:1999qv}, and final 
states beyond pair production\cite{Dvergsnes:2002nc}. The general helicity 
formalism for spin-two particles has been developed in 
Ref.~\cite{Gleisberg:2003ue}.

Constraints have also been placed on these theories of extra dimensions by 
testing the gravitational inverse-square law. The case of $n=1$ is already 
ruled out by solar system observations, and tests at the sub-millimeter 
level\cite{Hoyle:2004cw} can provide bounds at the TeV level (and
hence comparable to bounds obtained in collider experiments like the
one discussed in this paper) for $n=2$.

\section{Helicity Amplitudes for $\boldmath{\gamma \gamma \to \gamma \gamma}$}

Feynman rules have been developed for the KK
compactification of $n$ extra 
dimensions on a torus $T^n$ with all of the $n$ compactification radii 
equal\cite{Han:1998sg}. Using the couplings of the $d=4$ gauge fields to
gravity, one can analyze the possible effects of low scale gravity on 
gauge boson scattering. Since this phenomenology involves the exchange of 
massive spin-two KK states, there is a possibility of 
unique angular dependences
in cross sections involving the exchange of these quanta. 

We define momentum and polarization vectors for the initial 
and final particles 
as\footnote{This choice of polarization vectors is the same as the 
one in Ref.~\cite{Jikia:1993di}. Our definitions of the Mandelstam variables
require switching $t$ and $u$ when comparing with that paper.}
\begin{eqnarray}
p_{1} = \frac{\sqrt{s}}{2} (1 ; 0, 0, 1)
&&
\qquad p_{2} = \frac{\sqrt{s}}{2} (1 ; 0, 0, -1)\nonumber\\
\nonumber \\
k_{1} = \frac{\sqrt{s}}{2} (1 ; \beta\sin\theta, 0, \beta\cos\theta)
&&
\qquad k_{2} = \frac{\sqrt{s}}{2} 
(1 ; -\beta\sin\theta, 0, -\beta\cos\theta)\nonumber\\
\nonumber \\
e^{+}_{1}=e^{-}_{2}=\frac{1}{\sqrt{2}}(0;-1,-i,0) 
&& \qquad
e^{-}_{1}=e^{+}_{2}=\frac{1}{\sqrt{2}}(0;1,-i,0)\nonumber
\end{eqnarray}
\begin{eqnarray}
e^{+*}_{3}=e^{-*}_{4}=\frac{1}{\sqrt{2}}(0;-\cos\theta,i,\sin\theta)
\nonumber
\end{eqnarray}
\begin{eqnarray}
e^{-*}_{3}=e^{+*}_{4}=\frac{1}{\sqrt{2}}(0;\cos\theta,i,-\sin\theta)
\nonumber
\end{eqnarray}
\begin{eqnarray}
e^{0}_{3}=\frac{\sqrt{s}}{2m_{z}}(\beta;\sin\theta,0,\cos\theta)
\nonumber
\end{eqnarray}
\begin{eqnarray}
e^{0}_{4}=\frac{\sqrt{s}}{2m_{z}}(\beta;-\sin\theta,0,-\cos\theta)
\nonumber
\end{eqnarray}
\\
where $\beta = 1$ for the $\gamma\gamma\rightarrow\gamma\gamma$ case and 
$\beta = \sqrt{1-{{4M_Z^2}\over {s}}}$ 
for the $\gamma\gamma\rightarrow$ $ZZ$ case, 
$s = (p_{1} + p_{2})^{2}$~,  $t = (p_{1} - k_{1})^{2}$, 
and $u = (p_{1} - k_{2})^{2}$.\\  

The process $\gamma\gamma\rightarrow\gamma\gamma$ can be expressed in 
terms of three independent helicity amplitudes. The other helicity amplitudes
are related to these three by virtue of crossing relations and parity 
considerations. For the graviton 
exchange signal we find that only two of these 
three are nonvanishing, 
\begin{eqnarray}
i{\mathcal M}_{++++}^{\ga\ga} &=& -\frac{\kappa^{2}}{2}\Big(D_{E}(t) 
+ D_{E}(u)\Big)s^{2}\nonumber \;, \\
&=&-\frac{\kappa^{2}}{2}\bigg(D_{E}\Big(-\frac{s}{2}(1-\cos\theta)\Big)
+ D_{E}\Big(-\frac{s}{2}(1+\cos\theta)\Big)\bigg)s^{2}\nonumber \;, \\
i{\mathcal M}_{++--}^{\ga\ga} &=&-\frac{\kappa^{2}}{4}\Big(D_{E}(t)
-D_{E}(u)\Big)\Big(u^{2}-t^{2}\Big)\nonumber\\
&=&-\frac{\kappa^{2}}{4}\bigg(D_{E}\Big(-\frac{s}{2}(1-\cos\theta)\Big)-D_{E}
\Big(-\frac{s}{2}(1+\cos\theta)\Big)\bigg)s^{2}\cos\theta\nonumber \;, \\
i\mathcal{M}_{+++-}^{\ga\ga} &=&0
\end{eqnarray}
where $D(x)$ for $x=s$ and $D_{E}(x)$ for $x=t,u$ 
are the summed propagator 
functions\footnote{We find the sometimes used approximations 
\begin{eqnarray}
\kappa ^2D(s)\approx -{{16\pi i}\over {M_S^4}}{\mathcal F}\;,\label{aprx}
\end{eqnarray}
where 
\begin{eqnarray}
{\mathcal F}=\left \{\begin{array}{lr}
\log\left ({{M_S^2}\over s}\right )& {\mathrm for} \:\: n=2\\
& \\
{2\over {n-2}} &
{\mathrm for} \:\:n>2\;.
\end{array} \right .
\end{eqnarray}
can cause deviations from the exact expressions of tens of percent in the 
cross section.} 
derived in Ref.~\cite{Han:1998sg} and $\kappa =\sqrt{16\pi G_N}$. 
We have therefore
used the full expression for $D(s)$ for our analysis
of the $\gamma \gamma \to ZZ$ process, which is
\begin{eqnarray}
D(s) &=& {{s^{{n\over 2}-1}R^n}\over 
{(4\pi)^{n/2}\Gamma(\frac{n}{2})}}
\left (\pi+2iI\left (\frac{M_{S}}{\sqrt{s}}
\right )\right )\;, \label{Ds}
\end{eqnarray}
where
\begin{eqnarray}
I(x)&=&\left\{
\begin{array}{lr}
-\sum_{k=1}^{{n\over 2}-1}{1\over {2k}}x^{2k}
-{1\over 2}\log (x^2-1)& \qquad n=\mathrm{even}\\
& \\
-\sum_{k=1}^{{{n-1}\over 2}-1}{{1}\over {2k-1}}
x^{2k-1}
+{1\over 2}\log\left ({{x+1}\over {x-1}}\right )& 
\qquad n=\mathrm{odd}
\end{array}
\right. \;,
\end{eqnarray}
and
\begin{eqnarray}
D_{E}(t) &=& {{|t|^{{n\over 2}-1}R^n}\over
{(4\pi)^{n/2}\Gamma({n\over 2})}}
\left (-2iI_E\left (
\frac{M_{S}}{\sqrt{|t|}}\right )\right )\;,
\end{eqnarray}
where
\begin{eqnarray}
I_E(x)&=&\left\{
\begin{array}{cc}
(-1)^{{n\over 2}+1}\sum_{k=1}^{{n\over 2}-1}{{(-1)^{k}}\over {2k}}
x^{2k}
+{1\over 2}\log (x^2+1)& \qquad n=\mathrm{even}\\
& \\
(-1)^{{n-1}\over 2}\sum_{k=1}^{{n-1}\over 2}{{(-1)^{k}}\over {2k-1}}
x^{2k-1}
+\tan^{-1}(x)& \qquad n=\mathrm{odd}
\end{array}
\right. \;.
\end{eqnarray}
The scale $M_S$ is defined as 
\begin{eqnarray}
R^n={{(4\pi)^{n/2}\Gamma(n/2)}\over {2M_S^{n+2}G_N}}\;,
\end{eqnarray}
where $G_N=1/(8\pi\bar{M}_{pl}^2)$ is the 4-dimensional Newton's constant, with
$\bar{M}_{pl}=2.4\times 10^{18}$~GeV is the reduced Planck mass. 
This definition for the mass scale $M_S$
is the one of Han, Lykken, and Zhang\cite{Han:1998sg} and makes precise the
relationship in Eq.~\ref{scales}.
Other possible conventions for the mass scale were considered in 
Refs.~\cite{Giudice:1998ck,Hewett:1998sn} and 
should not be confused with the one chosen here.

The amplitudes ${\mathcal M}_{++++}^{\ga\ga}$ and 
${\mathcal M}_{+-+-}^{\ga\ga}$ and also 
${\mathcal M}_{++++}^{\ga\ga}$ and ${\mathcal M}_{+--+}^{\ga\ga}$ are related
by crossing 
\begin{eqnarray}
{\mathcal M}_{+--+}^{\ga\ga} (s,t,u)&=& {\mathcal M}_{++++}^{\ga\ga}(u,t,s)
\nonumber \;, \\
{\mathcal M}_{+-+-}^{\ga\ga} (s,t,u) &=& {\mathcal M}_{++++}^{\ga\ga}(t,s,u)
\;.
\end{eqnarray}
It is also noteworthy that the matrix element ${\mathcal M}_{++--}$
vanishes in the approximation $D(s)\approx D_E(|t|)\approx D_E(|u|)$.

Representing the initial and final helicity 
states of the photons as $\lambda_{1}\lambda_{2}$
and $\lambda_{3}\lambda_{4}$, respectively, the other four non-zero helicity 
amplitudes can be expressed in terms of one of the previous amplitudes through
\begin{eqnarray}
{\mathcal M}_{\lambda_{1}\lambda_{2}\lambda_{3}\lambda_{4}}^{\ga\ga}(s,t,u)&=
&{\mathcal M}_{-\lambda_{1}-\lambda_{2}-\lambda_{3}-\lambda_{4}}^{\ga\ga}
(s,t,u)\;,
\label{relation1}\\
{\mathcal M}_{\lambda_{1}\lambda_{2}\lambda_{3}\lambda_{4}}^{\ga\ga}(s,t,u)&=
&{\mathcal M}_{\lambda_{2}\lambda_{1}\lambda_{4}\lambda_{3}}^{\ga\ga}(s,t,u)
\;, 
\label{relation2}
\end{eqnarray}
which are results of Bose symmetry and parity. The angular dependence of 
these matrix elements are in agreement
with the results of Ref.~\cite{Atwood:1999cy}.

By squaring and summing these matrix elements and making the approximation as 
in Eq.~(\ref{aprx}), the polarization averaged result for the signal 
only (without the 
Standard Model background) can be derived, 
namely
\begin{eqnarray}
{1\over 4}\sum|{\mathcal M}^{\ga\ga}|^2
={{\kappa ^4}\over 2}|D(s)|^2(s^4+t^4+u^4)\;.\label{2phosig}
\end{eqnarray}
The factor ${1\over 4}$ is the 
initial state photon polarization average.
This is in agreement with the corresponding result  
in Ref.~\cite{Cheung:1999ja}
if an erroneous factor of one-half in the KK propagator of an earlier version 
of Ref.~\cite{Han:1998sg} is omitted. Furthermore, our result agrees with 
Ref.~\cite{Cheung:1999ja}'s expression when written in terms of $M_{S}$.

The signal represented by these amplitudes for 
$\gamma \gamma \to \gamma \gamma$ at photon-photon colliders has been studied 
before\cite{Cheung:1999ja,Davoudiasl:1999di,Choudhury:1999gp}. 
Explicit analytic
expressions for the helicity amplitudes allow one to understand more fully
the optimal strategy for exploiting polarization to optimize the sensitivity.
In our discussion
of the process $\gamma \gamma \to ZZ$ beginning in the next section,
we will be able to compare to the simpler case 
of $\gamma \gamma \to \gamma \gamma$ and highlight 
some important contrasts. A detailed analysis
of $\gamma \gamma \to \gamma \gamma$ as a mode to study exchange of KK
states at photon-photon colliders will appear elsewhere\cite{BKZ}.

\section{Helicity Amplitudes for $\gamma \gamma \to ZZ$}

The graviton exchange Feynmann rules for the $\gamma\gamma\rightarrow Z Z$ 
process is similar to the 
$\gamma\gamma\rightarrow\gamma\gamma$ case except for the restriction of 
the process to the s-channel.
This restriciton is due to the fact that there 
is no interaction vertex between 
$\gamma$, $Z$, and the KK state.
We define
\begin{eqnarray}
s_{4} &=& s - 4M_Z^2 \nonumber \;, \\
Y &=& tu - M_Z^4 = s\cdot p^{2}_{TZ}\;,
\end{eqnarray}
where $p_{TZ}^{}$ is the transverse 
momentum of either $Z$.
For the TT polarization modes (the notation T  
denotes collectively the two transverse polarizations ($+$ and $-$) of the 
$Z$ boson and L will denote the longitudinal polarization ($0$)).
for the final state $Z$ bosons, we obtained\footnote{We have chosen to denote
the helicity amplitudes for $\ga\ga\to\ga\ga$ by ${\mathcal M}^{\ga\ga}$ and
those for our main focus $\ga\ga\to ZZ$ without any superscript.}
\begin{eqnarray}
i{\mathcal M}_{+-++}=i{\mathcal M}_{+---}&=&-D(s)2\kappa^{2}
{{Y}\over {s_{4}}}M_Z^2 
\nonumber \\
&=&-D(s){{\kappa^{2}M_Z^2s}\over {2}}\sin^{2}\theta \;, \\
i{\mathcal M}_{+--+}&=&D(s){{\kappa^{2}}\over {4\beta^3}}\left (2\beta M_Z^4
-2(t-u)M_Z^2-t^2(1+\beta)+u^2(1-\beta)\right ) 
\nonumber \\
&=&-D(s){{\kappa^{2}s^{2}}\over {8}}(1-\cos\theta)^{2} \;, \\
i{\mathcal M}_{+-+-}&=&D(s){{\kappa^{2}}\over {4\beta^3}}\left (2\beta M_Z^4
-2(u-t)M_Z^2-u^2(1+\beta)+t^2(1-\beta)\right )   
\nonumber \\
&=&-D(s){{\kappa^{2}s^{2}}\over {8}}(1+\cos\theta)^{2} \;.
\end{eqnarray}
The amplitudes ${\mathcal M}_{+-+-}$ and ${\mathcal M}_{+--+}$ are related
either by $t\leftrightarrow u$ or by $\beta \to -\beta$.
For the LL final state polarization mode, we obtained
\begin{eqnarray}
i{\mathcal M}_{+-00}&=&D(s){{\kappa^{2}Y}\over {2s_{4}}}(s+4M_Z^2) 
\nonumber \\
&=&D(s){{\kappa^{2}s}\over {8}}(s+4M_Z^2)\sin^{2}\theta \;.
\end{eqnarray}
Finally for the TL final state polarization modes, we obtained
\begin{eqnarray}
i{\mathcal M}_{+-+0}=-i{\mathcal M}_{+-0-}&=&
-D(s){{\kappa^{2}\Delta Y}\over {\beta^2}}
\left (\beta+{{t-u}\over {s}}\right ) 
\nonumber \\
&=&-D(s){{\kappa^{2}M_Zs}\over {2}}
\sqrt{{{s}\over {2}}}\sin\theta(1+\cos\theta) \;, \\
i{\mathcal M}_{+--0}=-i{\mathcal M}_{+-0+}&=&
-D(s){{\kappa^{2}\Delta Y}\over {\beta^2}}
\left (\beta+{{u-t}\over {s}}\right ) 
\nonumber \\
&=&-D(s){{\kappa^{2}M_Zs}\over {2}}
\sqrt{{{s}\over {2}}}\sin\theta(1-\cos\theta) \;,
\end{eqnarray}
with $\Delta = \sqrt{{sM_Z^2}\over {2Y}}$.
Other helicity modes can be obtained from these by using equations analogous
to Eqns.~(\ref{relation1})-(\ref{relation2}). The first of these equations 
must be modified to account for the possibility of the TL final state
\begin{eqnarray}
{\mathcal M}_{\lambda_{1}\lambda_{2}\lambda_{3}\lambda_{4}}(s,t,u,\beta)&=
&{\mathcal M}_{-\lambda_{1}-\lambda_{2}-\lambda_{3}-\lambda_{4}}(s,t,u,\beta)
(-1)^{\lambda _3-\lambda_4}\;, \label{relation3} \\
{\mathcal M}_{\lambda_{1}\lambda_{2}\lambda_{3}\lambda_{4}}(s,t,u,\beta)&=
&{\mathcal M}_{\lambda_{2}\lambda_{1}\lambda_{4}\lambda_{3}}(s,t,u,\beta)
\;.
\label{relation4}
\end{eqnarray}
This amounts to an extra minus sign only. One can also obtain a relation 
between TL amplitudes that amounts to taking $\beta \to -\beta$, but we have
chosen to display these helicity amplitudes separately to emphasize their 
relationship under the interchange $t\leftrightarrow u$.

Helicity modes ${\mathcal M}_{-+\lambda_3\lambda_4}$ can be obtained from 
the corresponding amplitudes ${\mathcal M}_{+-\lambda_3\lambda_4}$
All other independent helicity amplitudes vanish; in particular, the signal
vanishes if the initial photons have the same helicity.
We again find agreement with the angular dependence of these 
helicity amplitudes with those in Ref.~\cite{Atwood:1999cy}.

At high energies ($\sqrt{s}>>M_Z$) the Standard Model background 
is dominated by $Z$ bosons in the  
transverse polarization states. There are contributions from all initial
helicity possibilities of the incident photons. 
The Higgs boson contributes only to channels in which the two initial photons
have the same helicity ($\lambda_{1}=\lambda_{2}$) and the 
final states $Z$ bosons must 
have the same helicities ($\lambda_{3}=\lambda_{4}$). 
This property reflects the fact that the Higgs boson is spin-zero,
and while the Higgs boson does not appreciably affect the results for the 
low scale gravity signal, we mention it here to contrast it with the 
spin-two nature of the $s$-channel graviton exchange graphs. 

The $s$-channel graviton exchange graphs 
require differing helicities ($\lambda_{1}=-\lambda_{2}$) for the 
initial photons. The 
dominant matrix elements for high energies ($\sqrt{s}>>M_Z$) are 
${\mathcal M}_{+-+-}$, ${\mathcal M}_{+--+}$ and ${\mathcal M}_{+-00}$ which 
have the following angular dependences respectively: 
$t^2={s^2\over 4}(1-\cos\theta)^2$, 
$u^2={s^2\over 4}(1+\cos\theta)^2$, and $tu={s^2\over 4}\sin^2\theta$.
The absence of a signal in channels where the initial photons have the same 
helicity differs from the $\gamma \gamma \to \gamma \gamma$ 
case, because in addition to the $s$-channel diagram, the 
$\gamma \gamma \to \gamma \gamma$ process
has additional contributions from the $t$ and $u$ channels. This
impacts the analysis in two ways: (1) For $\gamma \gamma \to ZZ$ 
one can try to isolate 
the signal by arranging the initial state helicities of the 
incoming photons to 
be opposite. This can be done by appropriately choosing the initial electron
and positron polarizations as well as the polarization of the backscattered 
laser beams. (2) The signal for $\gamma \gamma \to ZZ$ is somewhat 
smaller than the
signal for $\gamma \gamma \to \gamma \gamma$ expressed in 
Eq.~(\ref{2phosig}).  
This makes finding a signal harder, and 
weakens the overall bound one could otherwise place on the scale $M_S$.

Since the interference between the signal and the background can be crucial
to the detectability of any signal, it is important to examine not only their 
overall sizes but also their relative phases. At large energies, $s>>M_Z^2$, 
the Standard
Model background is dominated by the $W$ boson loops, and these dominant
contributions become predominantly imaginary\footnote{For explicit expressions,
see for example Eqn.~(3.26) of Ref.~\cite{Jikia:1993di}
or Eqn.~(10) of Ref.~\cite{Jikia:1993tc}.}. The signal involves the 
propagator function\cite{Han:1998sg}
\begin{eqnarray}
D(s)&=&\sum_{\vec{n}}{i\over {s-m_{\vec{n}}+i\epsilon}}\;.
\end{eqnarray}
Using 
\begin{eqnarray}
{1\over {s-m^2+i\epsilon}}=
P\left ({1\over {s-m^2}}\right )-i\pi\delta(s-m^2)\;,
\end{eqnarray}
yields the expression in Eq.~(\ref{Ds}), and one recognizes that the 
imaginary part of $D(s)$ contributes to the real part of the helicity
amplitudes, and the real part of $D(s)$ contributes to the imaginary part
of the helicity amplitudes. Physically speaking, the imaginary part of 
$D(s)$ involving $I(M_S/\sqrt{s})$ arises
from the (coherent) 
summation of the large number of nonresonant states and typically
dominates for $s<<M_S^2$. So in the physical region we are contemplating 
looking for a graviton exchange signal, $M_Z^2<<s<<M_S^2$, the background is
mostly imaginary and the signal is mostly real. One point that should not be
overlooked is that the approximation for $D(s)$ sometimes employed not only 
makes an approximation for the imaginary part, but also completely drops the
real part which can still have a significant interference with the 
$W$ loop Standard Model background. However it should be kept in 
mind that the $W$ loop background approaches its asymptotic behaviour
rather slowly, so the interference can still remain nonnegligible in practice
especially for the realistic case of $\sqrt{s_{ee}}=1$~TeV.

We find the TL polarization modes for the final state $Z$ bosons to be 
nonzero, but suppressed at high energies relative to the dominant helicity
amplitudes identified above by a factor $M_Z/\sqrt{s}$. These polarization 
modes are of course absent in the case of final state photons in 
$\gamma \gamma \to \gamma \gamma$. Finally the TT polarization modes
${\mathcal M}_{+-++}$ and ${\mathcal M}_{+---}$ are 
suppressed by a factor $M_Z^2/s$ because it requires
that the $Z$ bosons have the same helicity. This amplitude would vanish in 
the limit where $M_Z$ is taken to zero.

\section{Signal and Background}
Sources of high energy photons can be obtained by backscattering laser photons
of energy a few electron volts off high energy beams of electrons or 
positrons. Such 
colliders have come to be called photon-photon colliders or $\gamma \gamma$ 
colliders. This technique allows a much harder spectrum of photons than is 
available in the usual Weizs\"acker-Williams spectrum. In fact, photon-photon
collisions with energies almost the same order as the parent $e^+e^-$ collider
can be obtained. Furthermore, polarization of the electron and positron 
beams together with polarization of the lasers can yield polarized photon 
beams. Therefore by adjusting these polarizations, one can enhance
or suppress matrix elements with differing initial state photon helicities.
In the case of the $ZZ$ (and $W^+W^-$) final states, one can also in 
principle use the differing decay distributions to study the polarization 
states of the final state gauge bosons. This technique 
has not been employed in this analysis; we have imposed
instead a simple angular cut on the produced $Z$ bosons. 

The subprocess cross sections are given by 
$d\hat{\sigma}_{++}$ and $d\hat{\sigma}_{+-}$ where the final state 
polarizations have been summed over. Then the cross section folding in
the photon luminosity functions $f(x_i)$ and $\xi(x_i)$ for $i=1,2$, one 
obtains the differential cross section as
\begin{eqnarray}
d\sigma_{\lambda_3\lambda_4}&=&\int ^{y_m^2}_{M_Z^2/s_{ee}}d\tau
\int ^{y_m}_{\tau/y_m}{{dy}\over y}
f(y)f(\tau/y)\nonumber \\
&&\times \left [{1\over 2}\left \{1+\xi(y)\xi(\tau/y)\right \}
d\hat{\sigma}_{++\lambda_3\lambda _4}(s_{\gamma\gamma})
+{1\over 2}\left \{1-\xi(y)\xi(\tau/y)\right \}
d\hat{\sigma}_{+-\lambda _3\lambda_4}(s_{\gamma\gamma})\right ]\;,
\end{eqnarray}
where $y=E_\gamma/E_e$ and $\tau=s_{\gamma \gamma}/s_{ee}$ are the ratios 
of photon energies to the parent electron/positron energies.
The energy spectrum and helicity of backscattered photons, $f(y)$ and 
$\xi(y)$ are given in 
Refs.~\cite{Ginzburg:1981vm,Ginzburg:1982yr,Telnov:1989sd}.
We have taken the usual choice where the laser energy $\omega_0$ is chosen 
so that $x=4E_e\omega_0/m_e^2=2(1+\sqrt{2})\approx 4.8$ and 
$y_m=x/(x+1)\approx 0.83$.

The Standard Model background for $\gamma \gamma \to ZZ$ (and 
$\gamma \gamma \to \gamma \gamma$) is dominated by only 
a few helicity amplitudes at high energies.
For equal initial photon helicities the contribution to the cross section
from the amplitude ${\mathcal M}_{++++}$ is more than an order of magnitude 
larger than any other contribution even after a reasonable angular cut on the
final state $Z$ bosons. Similarly in the unequal initial photon helicity case 
the contribution to the cross section is dominated by the two amplitudes
${\mathcal M}_{+-+-}$ and ${\mathcal M}_{+--+}$. The cross section for 
longitudinally polarized $Z$ bosons arising from ${\mathcal M}_{+-00}$
is at least an order of magnitude smaller
for $\sqrt{s}_{\gamma \gamma}>500$~GeV.

These amplitudes are dominated at high energies by the $W$ loop contributions
(as opposed to the fermion loop diagrams), so the relative size of 
the cross sections for 
$\gamma \gamma \to \gamma \gamma$ and $\gamma \gamma \to ZZ$ is easily 
estimated in this limit. One simply substitutes for the relative sizes of the 
$\gamma \gamma W$ and $ZZW$ couplings: 
$\sigma(\gamma \gamma \to \gamma \gamma)=\tan ^4\theta_W
\sigma(\gamma \gamma \to ZZ)$, and the $ZZ$ final state is enhanced by a factor
of about twelve.

The fact that the signal for graviton exchange contributes only to 
helicity amplitudes with unequal initial photon helicities 
can be exploited experimentally. By selecting 
the electron, positron, and laser polarizations 
to give the desired initial photon 
helicities, one can suppress the large 
background arising from ${\mathcal M}_{++++}$ 
while enhancing the
signal. In contrast the process
$\gamma \gamma\to \gamma \gamma$ has signal contributions in both same and 
opposite initial photon helicity channels.  

We have assumed a Higgs boson mass of $M_H=150$~GeV to make the plots. 
A higher Higgs masses would appear as a resonance in some of the 
cross sections ($\sigma _{++00}$, $\sigma _{++++}$, and $\sigma _{++--}$),
but since the resonance is a small fraction of the background for any 
$M_H>400$~GeV, the exact value of the Higgs mass is completely irrelevant
for determining the size of the graviton 
signal plus background considered here. Similarly in the region where 
$\sqrt{s_{\ga\ga}}$ is several hundred TeV, the Standard Model $W$ loop
background completely dominates over the fermion loops. Nevertheless we
mention that we have used a top
quark mass of $m_t=175$~GeV, and occasionally one can notice a change in 
behaviour in the Standard Model background at the threshold 
$\sqrt{s_{\ga\ga}}=2m_t$.

The cross section for various helicity combinations of the initial state
photons and final state $Z$ bosons are shown in Figs. (1)-(5) for the Standard
Model background and for the graviton exchange
signal plus background for $n=4$ and for 
$M_S=3,4,5,6$~TeV. We have employed an angular cut on the c.o.m. scattering 
angle of $|\cos \theta | < \cos(\pi/6)$.
The signal is dominated by the cross sections 
$\sigma_{+-+-}$, $\sigma_{+--+}$, and $\sigma_{+-00}$ shown in 
Figs. (1) and (2).  For large 
$\sqrt{s_{\ga\ga}}$ ($\sqrt{s_{\ga\ga}}>>M_{z}$), the cross sections grow 
like $s_{\ga\ga}^3/M_S^8$. Moreover, for such large energy the 
$TL$ final state signals shown in Fig. (3) grow like $s_{\ga\ga}^2M_Z^2/M_S^8$ 
while the remaining $TT$ amplitudes shown in Fig. (4) grows like 
$s_{\ga\ga}M_Z^4/M_S^8$ for large 
$s_{\ga\ga}$. When the 
signal and background are of comparable size ($\sqrt{s_{\ga\ga}}\alt 1$~TeV),
the contribution from the transverse states 
shown in Fig. (1) will dominate the 
signal since the interference with the underlying Standard Model background 
determines its overall size. Therefore 
it is important to include the interference
between the signal and background when estimating the reach of possible 
future experimental searches.

The background consists of the Standard model contributions from the opposite
photon helicity ($\lambda_{1}=-\lambda_{2}$) modes shown in Figs. (1)-(4) as 
well as from the same photon helicity ($\lambda_{1}=\lambda_{2}$) modes shown 
in Fig. (5) for which, as previously 
mentioned, do not receive contributions from the spin-two graviton exchange.
The background is dominated by the cross section $\sigma _{++++}$ which 
can exceed 100~femtobarns. Unlike the process $\ga\ga\to\ga\ga$ there is no
signal contribution in this mode for $\ga\ga\to ZZ$ because the latter only 
proceeds via the $s$-channel. Furthermore the overall size of $\ga\ga\to ZZ$
is larger than $\ga\ga\to\ga\ga$ due to enhanced $WWZ$ coupling.
For most practical purposes the overall level of the signal and background
can be estimated by concentrating attention on the contributions in Figs. (1)
and (5) which dominate in most cases. We do 
not present a figure summing these contributions since the optimal strategy 
for uncovering the signal will be to use polarization to isolate the helicity 
amplitudes containing the signal as outlined in detail below.

\bigskip 
\centerline{\hbox{\epsfxsize=4.0in\epsffile{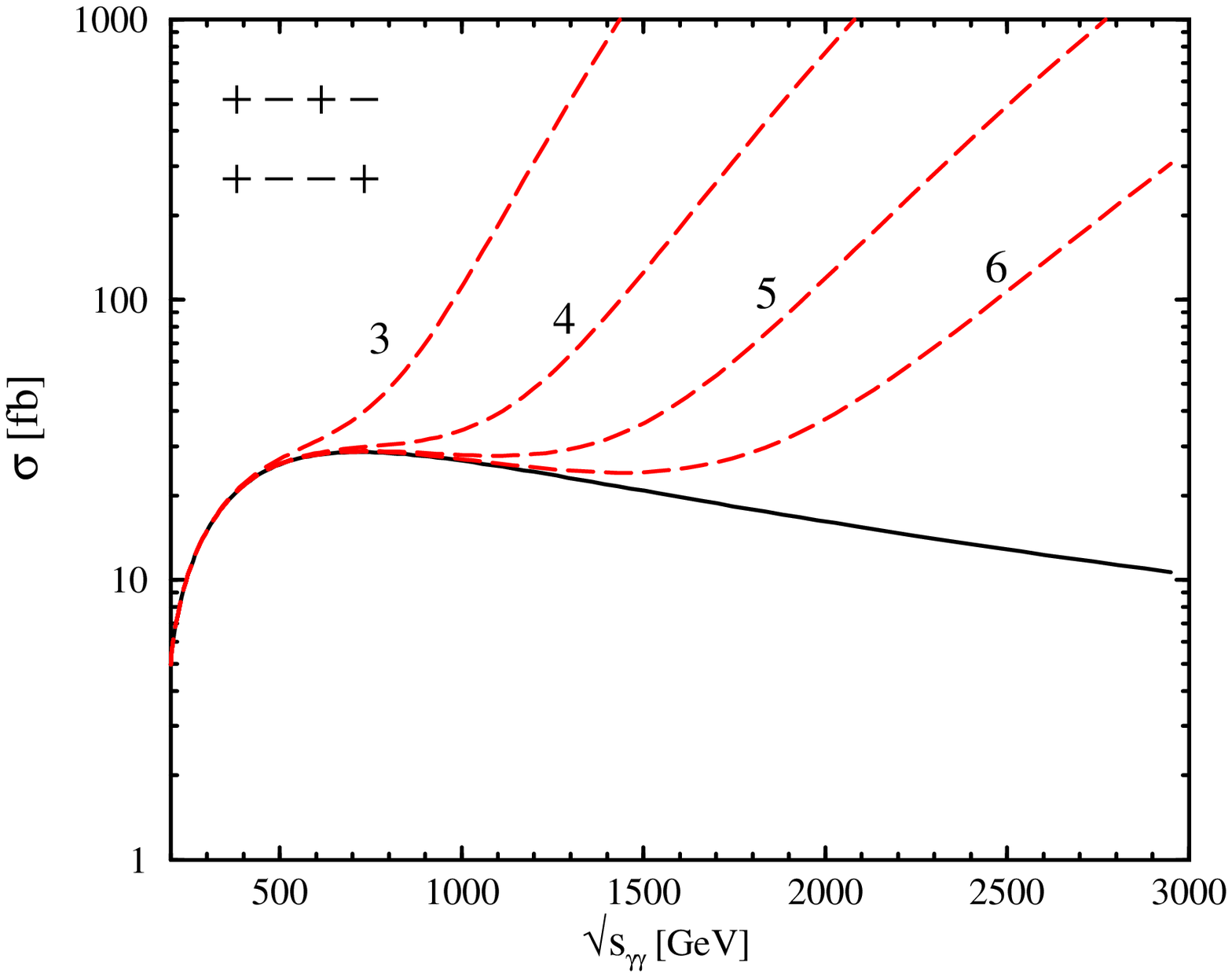}}}
\parbox{5.5in}{\small \noindent Fig. 1: The cross section is shown for
  $\sigma_{+-+-}=\sigma_{+--+}$ for the Standard Model background
  (solid) and for signal plus background (dashed) for $n=4$ and $M_S=3$~TeV,
  $4$~TeV, $5$~TeV, and $6$~TeV from top to bottom.  The signal cross
  sections grow like $s^3/M_S^8$ in the region $M_Z^2<<s<<M_S^2$. A
  cut has been placed on the c.o.m. scattering angle $|\cos
  \theta | < \cos(\pi/6)$.}  
\bigskip

\bigskip
\centerline{\hbox{\epsfxsize=4.0in\epsffile{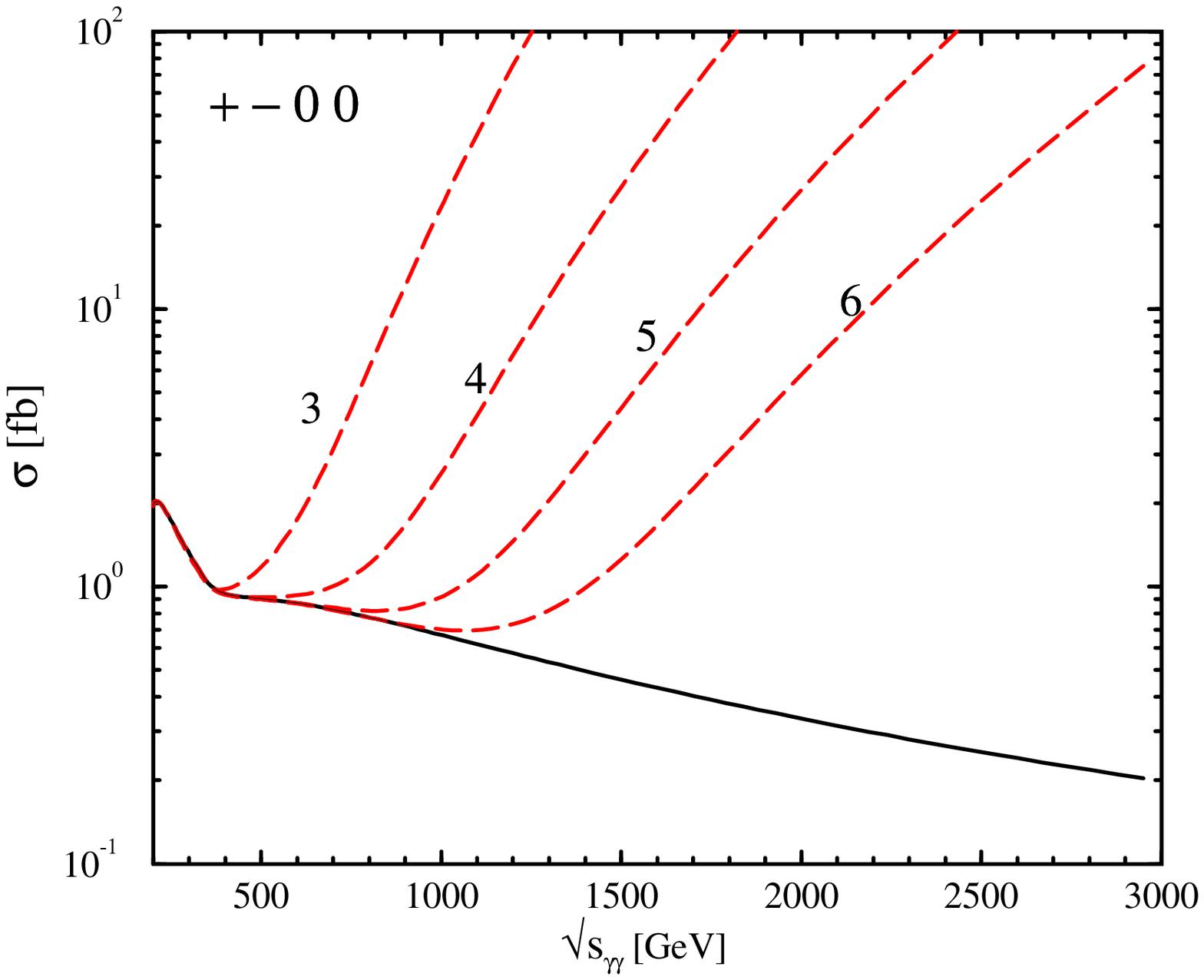}}}
\parbox{5.5in}{\small \noindent
Fig. 2: The cross section is shown for $\sigma_{+-00}$ for the Standard
Model background (solid) and for signal plus background (dashed)
for $n=4$ and $M_S=3$~TeV,
$4$~TeV, $5$~TeV, and $6$~TeV from top to bottom. The signal cross section 
grows like $s^3/M_S^8$ in the region $M_Z^2<<s<<M_S^2$. A cut has been placed 
on the c.o.m. scattering angle $|\cos \theta| < \cos(\pi/6)$.}
\bigskip

\bigskip
\centerline{\hbox{\epsfxsize=4.0in\epsffile{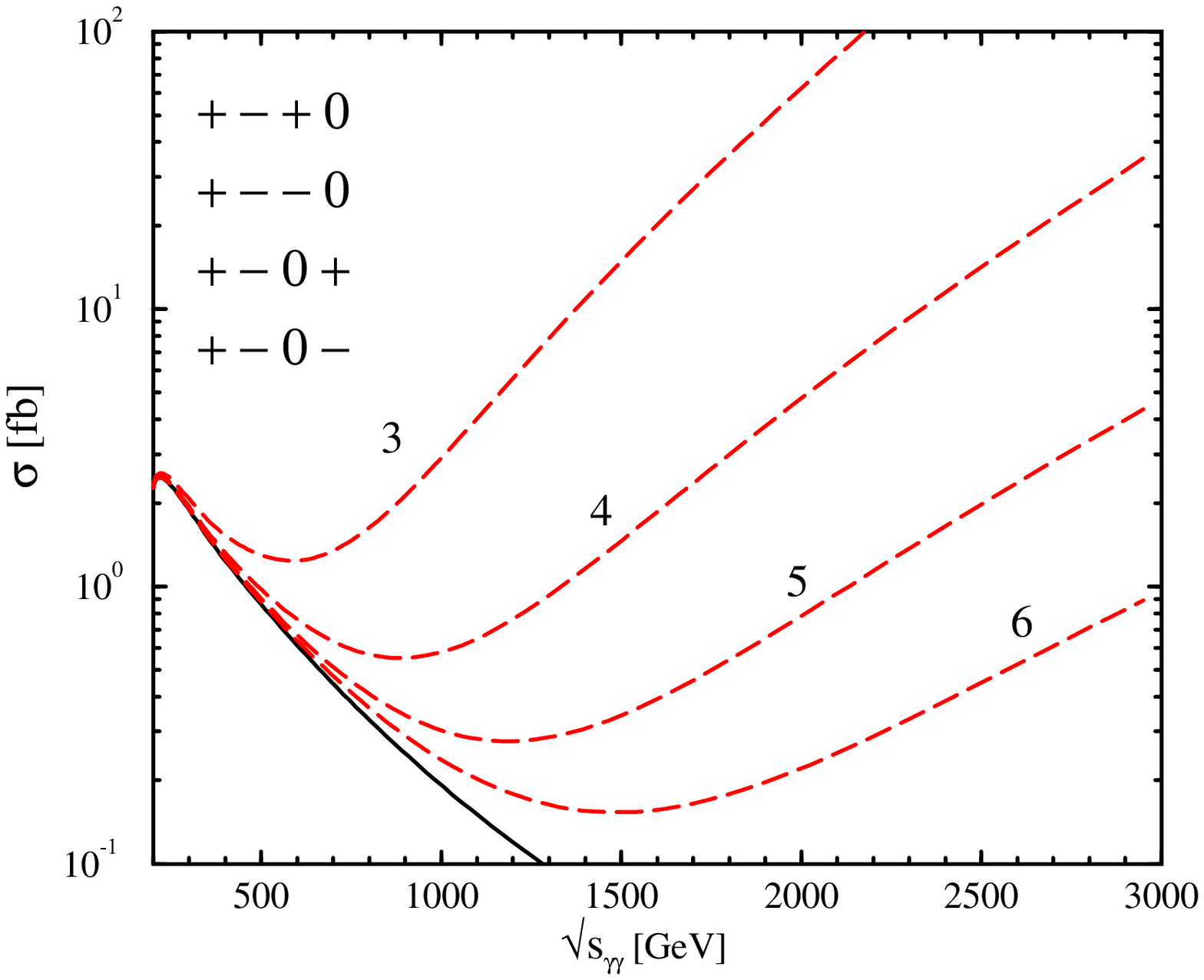}}}
\parbox{5.5in}{\small \noindent
Fig. 3: The cross section is shown for $\sigma_{+-+0}=\sigma_{+-0-}
(\approx \sigma_{+-0+}=\sigma_{+--0})$ 
for the Standard
Model background (solid) and for signal plus background (dashed)
for $n=4$ and $M_S=3$~TeV,
$4$~TeV, $5$~TeV, and $6$~TeV from top to bottom. The signal cross sections
grow like $s^2M_Z^2/M_S^8$ in the region $M_Z^2<<s<<M_S^2$. 
A cut has been placed 
on the c.o.m. scattering angle $|\cos \theta| < \cos(\pi/6)$.}
\bigskip

\bigskip
\centerline{\hbox{\epsfxsize=4.0in\epsffile{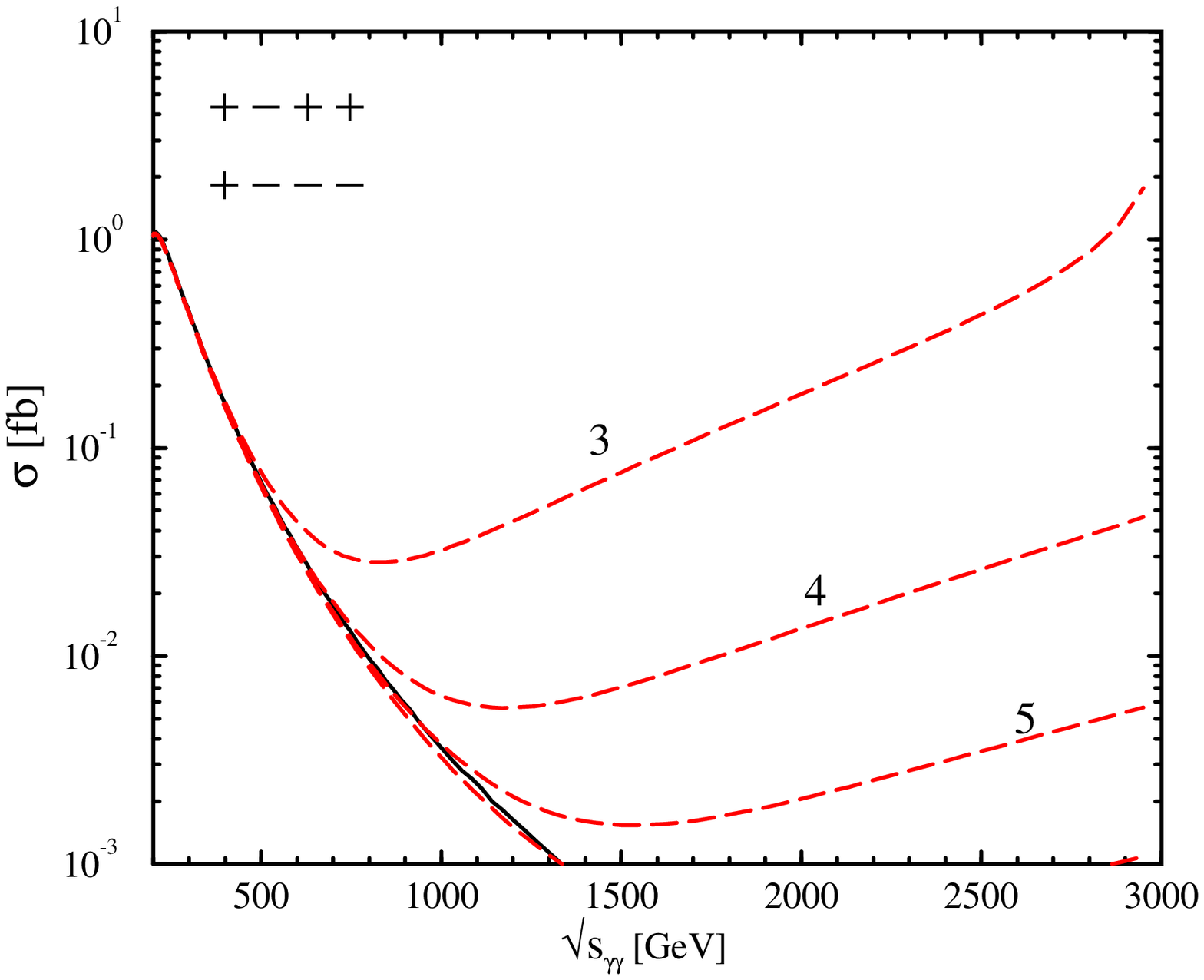}}}
\parbox{5.5in}{\small \noindent
Fig. 4: The cross section is shown for $\sigma_{+-++}=\sigma_{+---}$ 
for the Standard
Model background (solid) and for signal plus background (dashed)
for $n=4$ and $M_S=3$~TeV,
$4$~TeV, and $5$~TeV from top to bottom. The signal cross section grows 
like $sM_Z^4/M_S^8$ in the region $M_Z^2<<s<<M_S^2$. A cut has been placed 
on the c.o.m. scattering angle $|\cos \theta| < \cos(\pi/6)$.}
\bigskip

\bigskip
\centerline{\hbox{\epsfxsize=4.0in\epsffile{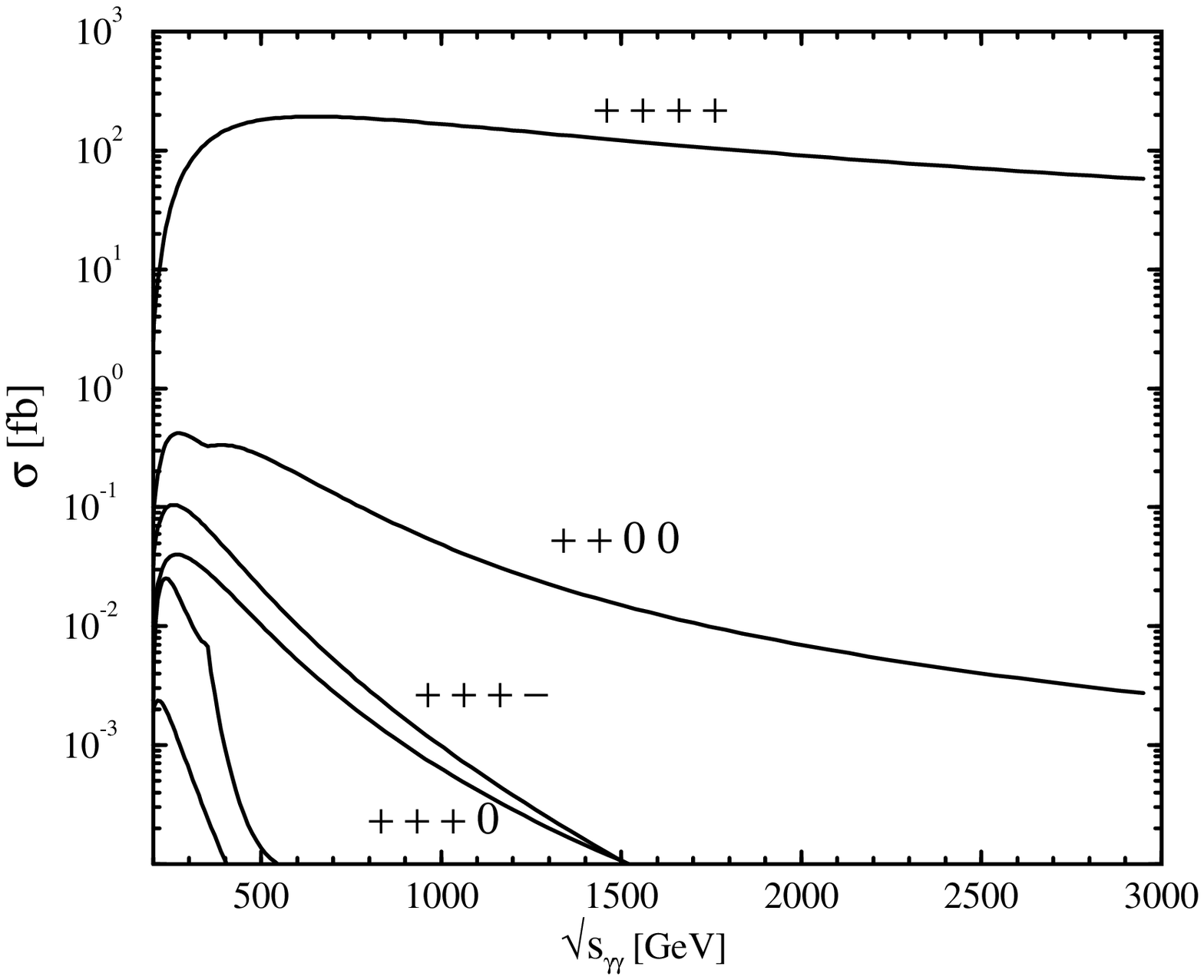}}}
\parbox{5.5in}{\small \noindent
Fig. 5: The cross sections for the case of equal photon helicities are shown.
Since the gravition signal does not contribute to these modes, what is shown
arises from the Standard Model alone and contributes only as background. Notice
the wide range of scales and the dominance of $\sigma_{++++}$ for the larger
$\sqrt{s_{\ga\ga}}$ of interest. At the lower left the unlabeled curves
correspond to $\sigma_{++--}$ (the larger one) and  $\sigma_{++-0}$ (the 
smaller one). A cut has been placed 
on the c.o.m. scattering angle $|\cos \theta| < \cos(\pi/6)$.}
\bigskip

\newpage

\bigskip 
\centerline{\hbox{\epsfxsize=4.0in\epsffile{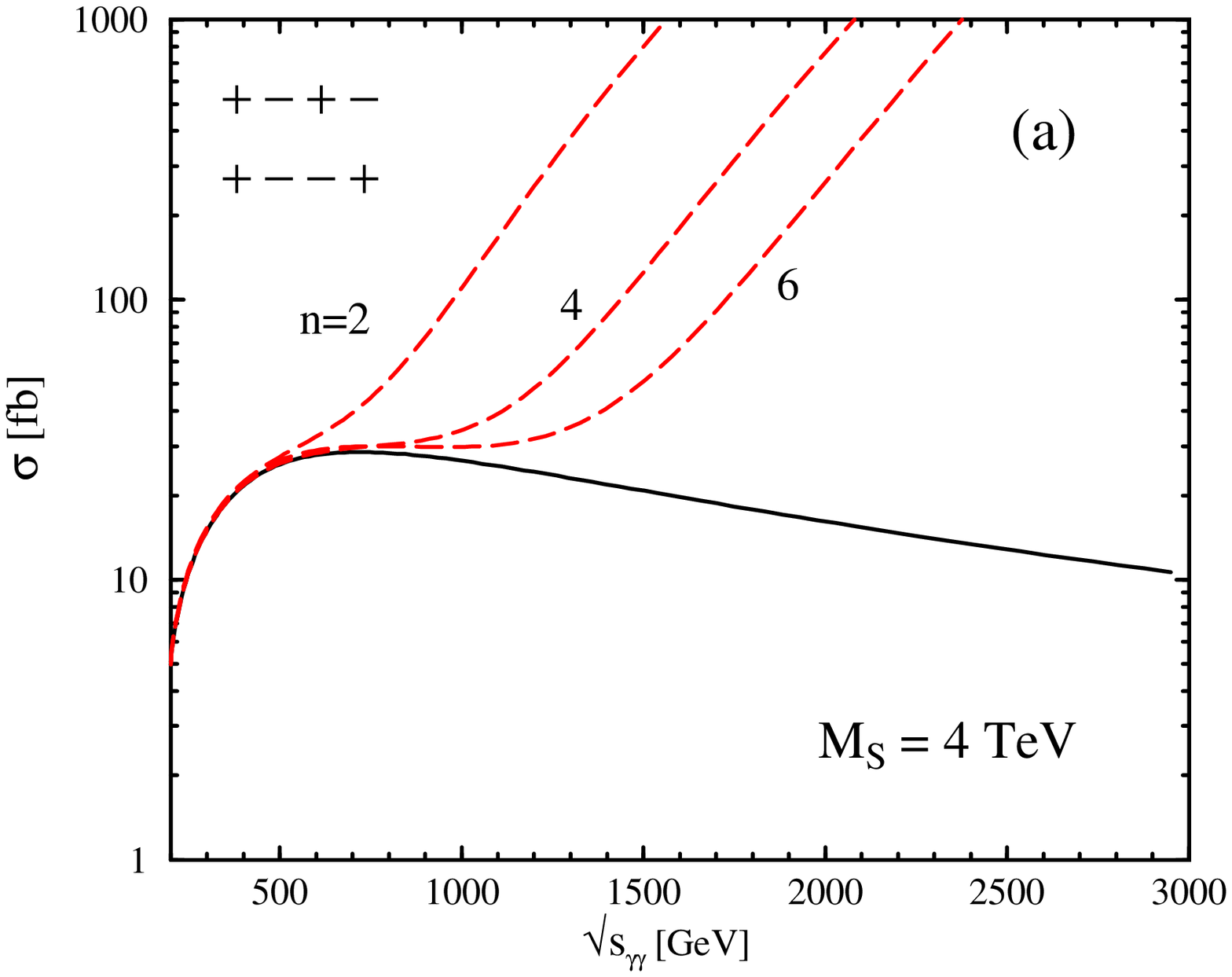}}}
\centerline{\hbox{\epsfxsize=4.0in\epsffile{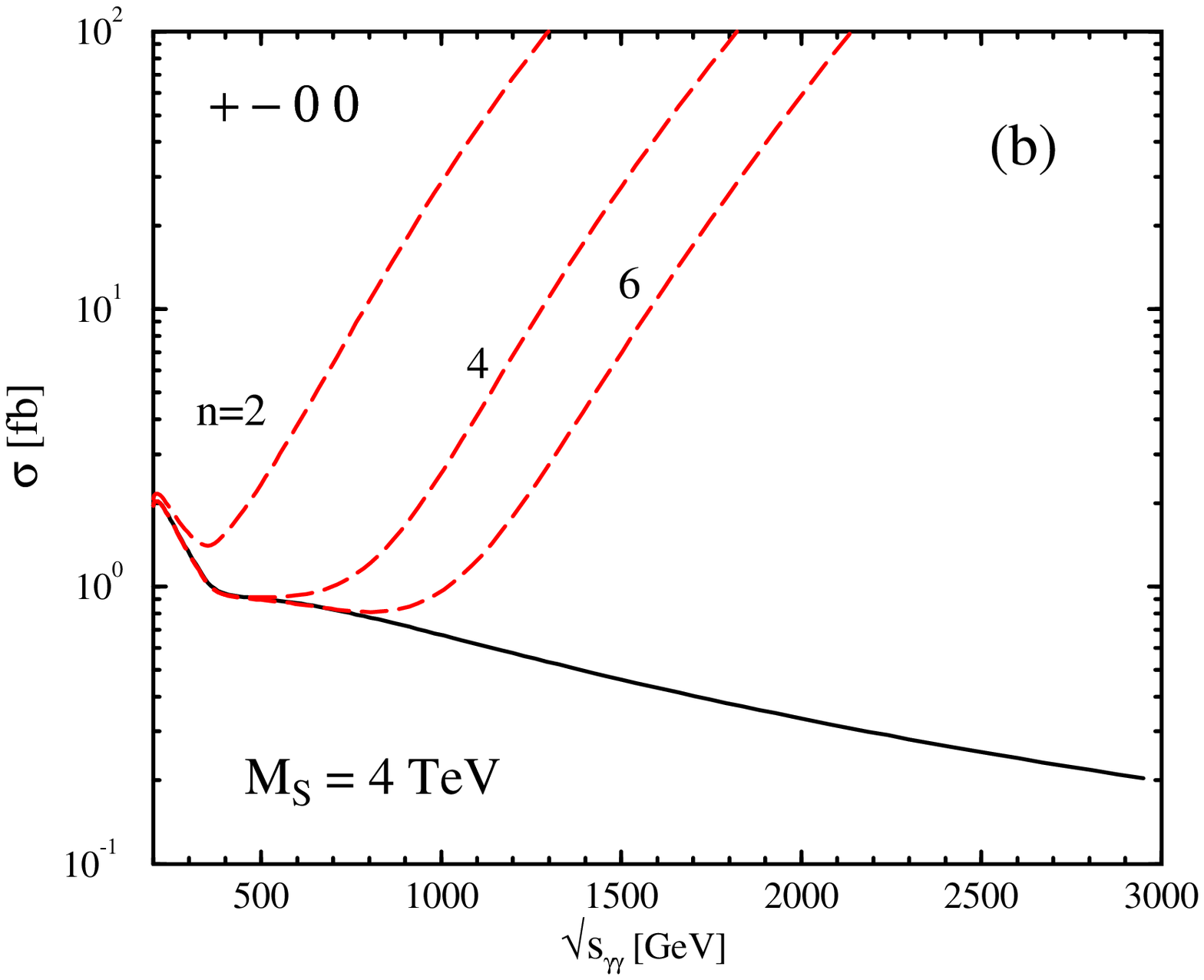}}}
\parbox{5.5in}{\small \noindent Fig. 6: The cross sections are shown for 
  (a) $\sigma_{+-+-}=\sigma_{+--+}$ and (b) $\sigma_{+-00}$
  for the Standard Model background
  (solid) and for signal plus background (dashed) for $M_S=4$~TeV and 
  the number of extra dimensions $n=2$, $4$, and $6$.}
\bigskip
\bigskip

In Fig. (6) the effect of varying the number of extra dimensions $n$ is 
shown keeping the scale $M_S$ fixed at 4~TeV. We show only the most 
important modes, namely $\sigma_{+-+-}=\sigma_{+--+}$ in Fig. (6a) and 
$\sigma_{+-00}$ in Fig. (6b). The conclusion is that stronger bounds can be
placed when $n$ is smaller.

The strategy of choosing polarizations to optimize the signal over background 
is particularly simple for the process $\ga\ga\to ZZ$.
The graviton exchange signal requires opposite helicities for the initial 
state photons, so one should choose polarizations for the electron and 
positron beams as well as the laser beams to isolate this combination and to 
eliminate as much as possible the large background from $\sigma_{++++}$.
We denote the polarizations of the electron ($e_1$), positron ($e_2$) and 
laser beams ($\ga_1$ and $\ga_2$) by 
$(P_{e_1},P_{\ga _1},P_{e_2},P_{\ga_2})$. 
At a photon-photon collider the luminosity is rather flat for 
the unpolarized case, and one achieves a peak in the luminosity just below
the maximum energy by choosing opposite polarizations for the electron
and laser photon, e.g. in the ideal case $P_{e_1}P_{\ga _1}=-1$ 
and $P_{e_2}P_{\ga _2}=-1$ (see for example Fig. (11) of 
Ref.~\cite{Telnov:1989sd}). Since one wants to look for a rapidly growing
signal on top of a Standard Model background, clearly the optimal situation 
occurs when the luminosities is concentrated at the highest energies possible.
In addition to isolate the opposite photon helicity amplitudes one wants to 
choose the polarizations such that $P_{e_1}=-P_{e_2}$ and 
$P_{\ga _1}=-P_{\ga _2}$. Therefore we have assumed in the following analysis
that the electron/positron beams can be polarized to 90\%, and assume the 
photon-photon collider has the following polarization combinations
\begin{eqnarray}
P_{e_1}&=&-P_{e_2}=0.9\;, \nonumber \\
P_{\ga _1}&=&-P_{\ga _2}=-1\;.
\end{eqnarray}
This polarization setting will be denoted by the shorthand 
$(P_{e_1},P_{\ga _1},P_{e_2},P_{\ga_2})=(+,-,-,+)$.
It was noticed in Ref.~\cite{Rizzo:1999sy} that this kind of polarization 
enhanced the signal for the process $\ga\ga\to W^+W^-$. This can be understood
on the basis of our helicity amplitudes for $\ga\ga\to ZZ$ which can be 
converted into helicity amplitudes for $\ga\ga\to W^+W^-$ with minor 
modifications since both processes occur via only the $s$ channel. The 
Standard Model background for $\ga\ga \to W^+W^-$ occurs at tree level rather
than at one-loop as it does for $\ga\ga\to ZZ$, so the reach is expected to 
be higher in $W$ production since the interference of the signal with the 
background is crucial.

The polarization setting that has the photon-photon luminosity peaking at 
the highest energy but gives predominantly backscattered photons with
the same helicity is $(P_{e_1},P_{\ga _1},P_{e_2},P_{\ga_2})=(+,-,+,-)$.
This polarization setting would be optimal for a case where a signal 
contributed to the helicity amplitudes ${\mathcal M}_{++\lambda_3\lambda_4}$.
Thus this setting would be preferable for the $\ga\ga\to\ga\ga$ process, which 
is consistent with the results of the calculations in 
Ref.~\cite{Davoudiasl:1999di}.

In Fig. (7) a comparison is made between the two polarization settings. 
One observes a noticeable improvement in the second polarization choice. 
For this choice we have determined the integrated luminosity required to 
observe at the 95\% confidence level a signal over the Standard Model 
background for three choices of $M_S$. This is shown in Fig. (8) for the
case of $n=4$. In particular, with an integrated luminosity of $100$~fb$^{-1}$,
a linear collider with c.o.m. energy of $1$~TeV has a reach almost up to 
$M_S=4$~TeV. This determination of the experimental reach  
for the case of  
$\ga\ga\to ZZ$ invites us to compare with the other diboson processes that have
been considered previously.
The reach is higher as expected for
$\ga\ga\to W^+W^-$ where the signal interferes with the much larger
tree-level background\cite{Rizzo:1999sy}. While a strategy of exploiting the 
decay products might favor the $ZZ$ final state with respect to the $W^+W^-$ 
final state, it will not be enough to overcome the different level of 
background. Of course, for high enough energies the signals become comparable
in size and the size of the backgrounds becomes irrelevant.  The reach in
$M_X$ is also slightly higher in $\ga\ga\to\ga\ga$
where contributions to the signal occur in the $t$ and $u$ channels as well as
the $s$ channel. This larger signal in $\ga\ga\to\ga\ga$ wins out against the 
larger level of Standard Model background in $\ga\ga\to ZZ$. In any event all
of these channels should be studied to determine the universality of the 
graviton couplings and to test whether the signal behaves as one expects from 
the exchange of a spin-two particle.


\bigskip 
\centerline{\hbox{\epsfxsize=4.0in\epsffile{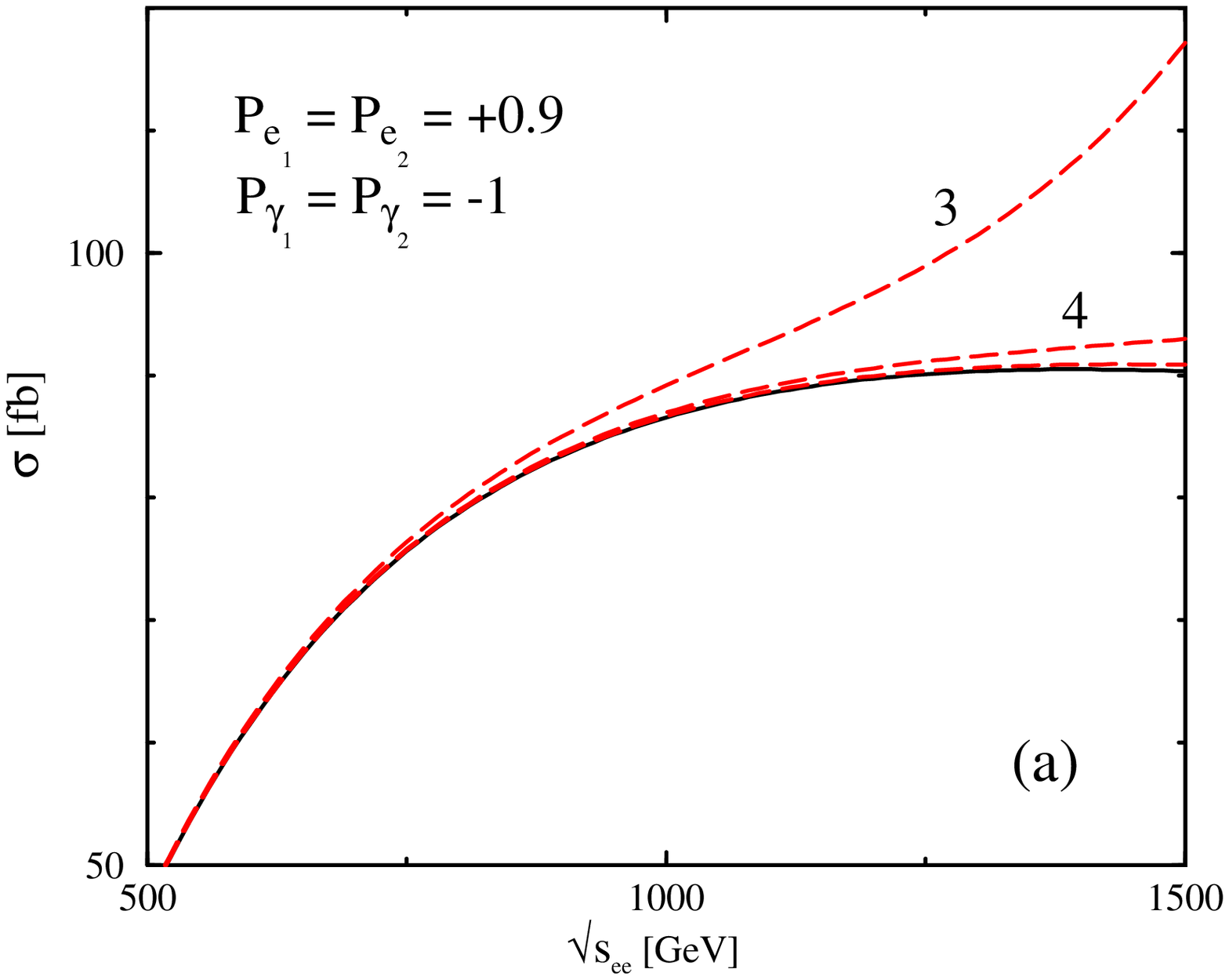}}}
\centerline{\hbox{\epsfxsize=4.0in\epsffile{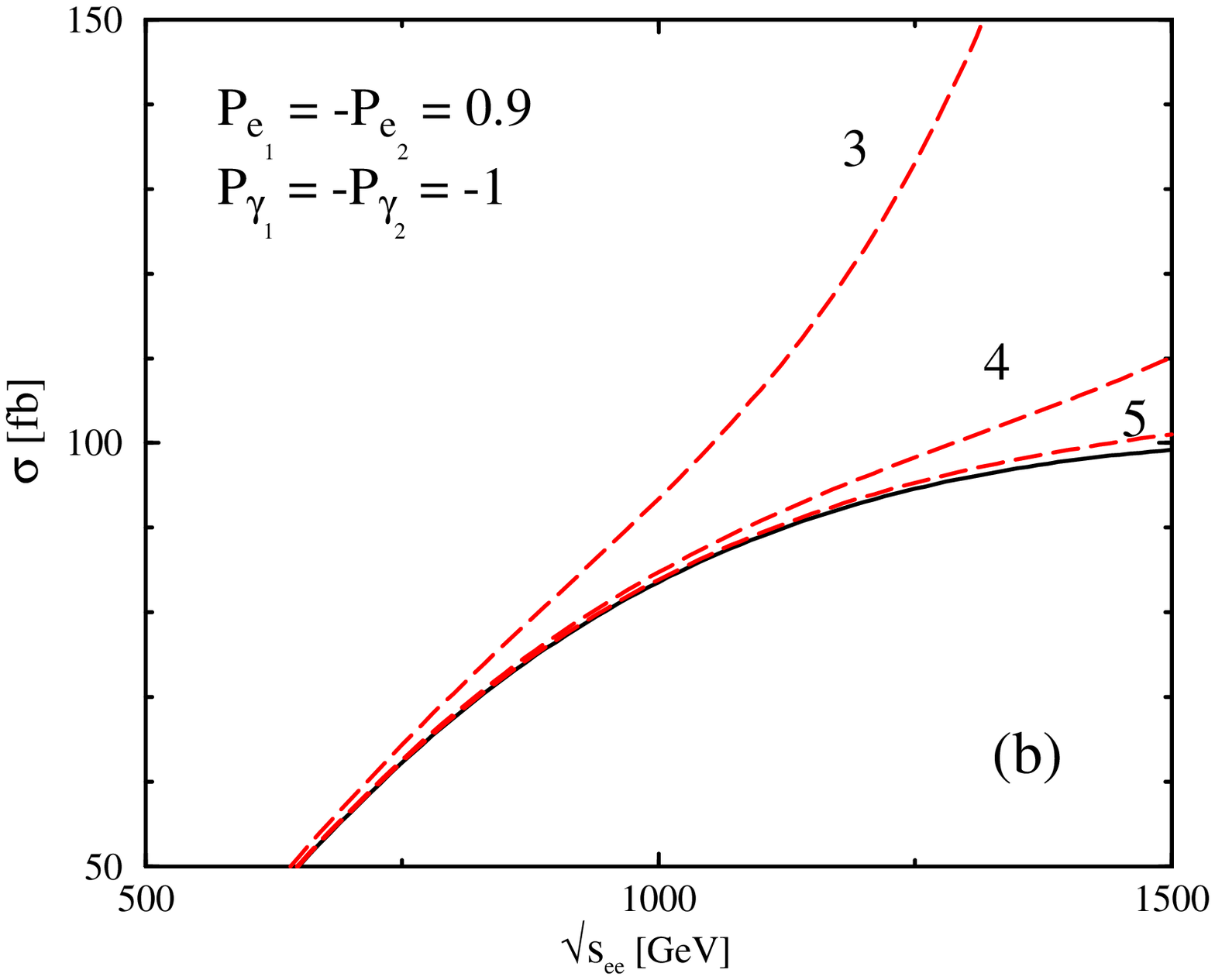}}}
\parbox{5.5in}{\small \noindent Fig. 7: The cross section are shown 
for a photon-photon
  collider whose parent $e^+e^-$ collider has energy $\sqrt{s_{ee}}$ for 
  the choice of polarizations 
(a) $(P_{e_1},P_{\ga _1},P_{e_2},P_{\ga_2})=(+,-,+,-)$
and 
(b) $(P_{e_1},P_{\ga _1},P_{e_2},P_{\ga_2})=(+,-,-,+)$,
and for $M_S=3,4,5$~TeV.
The number of extra dimensions is $n=4$.
The polarization in (a) favors backscattered photons with the same helicity 
while (b) favors backscattered photons with opposite helicities.}
\bigskip

\bigskip
\centerline{\hbox{\epsfxsize=4.0in\epsffile{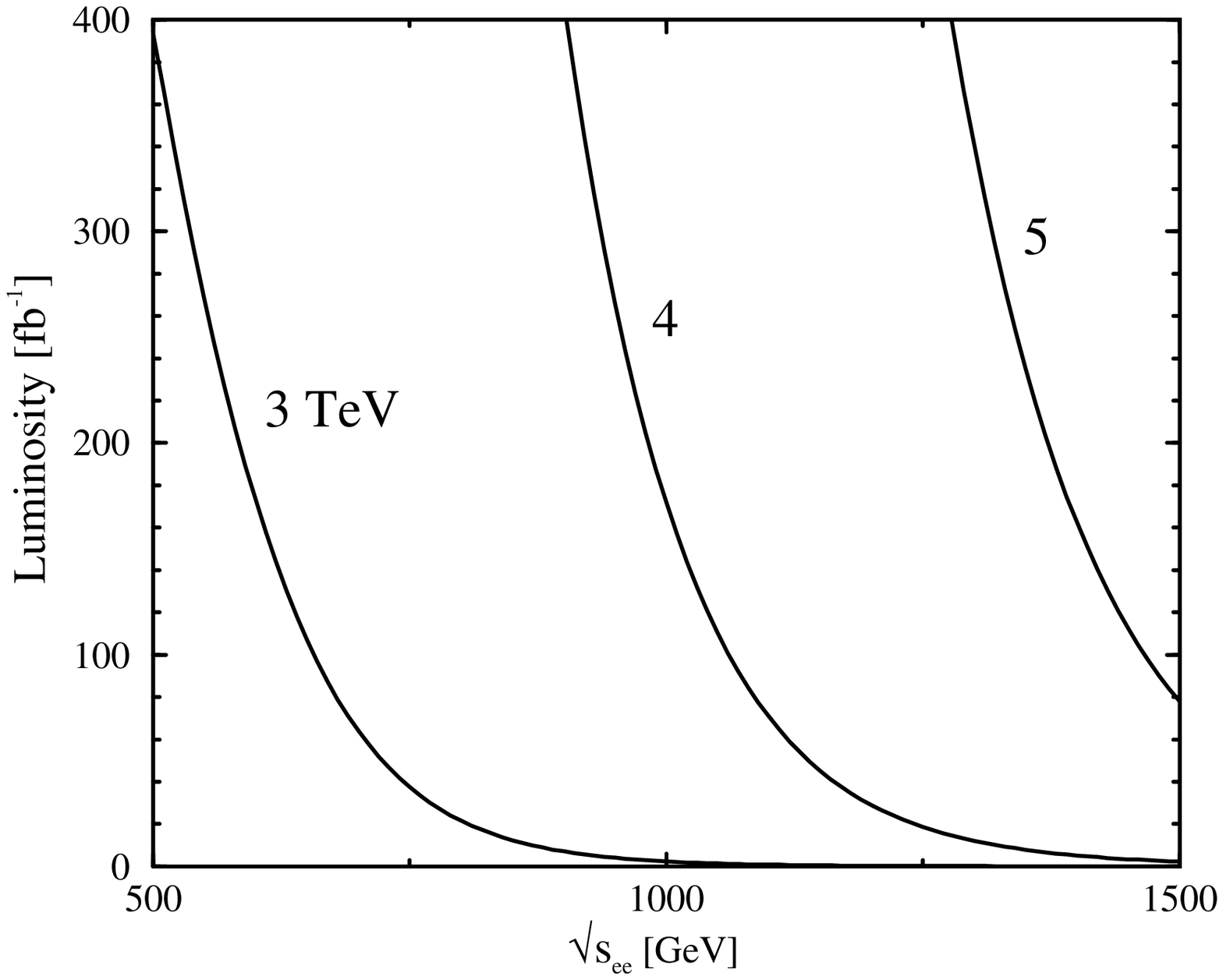}}}
\parbox{5.5in}{\small \noindent
Fig. 8: The luminosity required to detect required to detect a signal at 
the 95\% confidence level for $M_S=3,4,5$ TeV as a function of 
$\sqrt{s_{ee}}$ with the polarization choice
$(P_{e_1},P_{\ga _1},P_{e_2},P_{\ga_2})=(+,-,-,+)$ as in Fig. 7(b).
The number of extra dimensions in $n=4$.}
\bigskip


{\section{Conclusions} 
The processes $\gamma \gamma \to VV$ where $VV=ZZ$ or $W^+W^-$ are interesting 
reactions to look for any effects of low scale gravity. Unlike 
photon-photon scattering, $\gamma \gamma \to \gamma \gamma$, these cross
sections occur only via $s$-channel exchange of gravitons. Due to the spin-two 
nature of the exchanged quanta, this results in nonzero matrix elements only 
when the initial photons have opposite helicities. Exploiting the ability 
of Compton backscattering to provide a hard spectrum of polarized photons, 
one can hope to isolate a signal. 

We can suggest an overall strategy for analyzing all of the modes 
$\ga\ga\to VV$. Signals should be seen in all of the modes 
$\ga\ga\to ZZ$, $\ga\ga\to\ga\ga$, and $\ga\ga\to W^+W^-$ but should
be absent in $\ga\ga\to\ga Z$. The modes that occur only in the $s$ channel,
namely $\ga\ga\to ZZ$ and $\ga\ga\to W^+W^-$ should show a strong dependence
on the polarization settings of the photon-photon collider since only the
opposite helicity photons contribute to the signal. In particular
the polarization setting $(P_{e_1},P_{\ga _1},P_{e_2},P_{\ga_2})=(+,-,-,+)$
will enhance the signal by simultaneously resulting in opposite sign 
backscattered photon helicities and a peak in the photon-photon luminosity
at the highest energies. 
The signal-to-background
ratio $S/B$ for the photon-photon scattering process $\ga\ga\to\ga\ga$ should 
be less sensitive to the polarization setting. In this latter setting the 
polarizations to $(P_{e_1},P_{\ga _1},P_{e_2},P_{\ga_2})=(+,-,+,-)$ 
will enhance the 
sensitivity since the same photon helicity cross sections are larger than 
the opposite helicity cross sections.
 
If a graviton exchange is ever seen, then the angular dependences can 
be studied in detail. The rapid rise in the signal cross section means
that even modest enhancements in the photon-photon collider energy can 
yield dramatic improvements in the rates. 


\vspace{0.5cm}

\section*{Acknowledgments}

This work was supported in part by the U.S.
Department of Energy under Grant No.~DE-FG02-91ER40661.



\end{document}